\documentclass[conference]{IEEEtran}
\IEEEoverridecommandlockouts

\usepackage{cite}
\usepackage{amsmath,amssymb,amsfonts}
\usepackage{algorithmic}
\usepackage{graphicx}
\usepackage{textcomp}
\usepackage{xcolor}
\usepackage{tabularx,xcolor}
\usepackage{amssymb}

\usepackage{enumitem}  

\usepackage{amsmath}
\usepackage{multirow}
\usepackage{graphicx}
\usepackage{array}
\usepackage{float}
\usepackage{booktabs}
\usepackage{dblfloatfix}

\usepackage{setspace}

\def\BibTeX{{\rm B\kern-.05em{\sc i\kern-.025em b}\kern-.08em
    T\kern-.1667em\lower.7ex\hbox{E}\kern-.125emX}}
\begin{document}

\title{Vector Quantized Diffusion Model Based Speech Bandwidth  Extension\\
 
}

\author{
    Yuan Fang\textsuperscript{1,2,$\dagger$}, Jinglin Bai\textsuperscript{1,2,$\dagger$}, Jiajie Wang\textsuperscript{2}, Xueliang Zhang\textsuperscript{1} \\
    \textsuperscript{1}College of Computer Science, Inner Mongolia University, Hohhot, China \\
    \textsuperscript{2}SenseAuto Intelligent Connection, SenseTime, Beijing, China \\
    \{32209021,bjlin\}@mail.imu.edu.cn, wangjiajie1@senseauto.com, cszxl@imu.edu.cn
\thanks{$\dagger$ Work done during internship at SenseTime.}
}

\maketitle

\begin{abstract}
Recent advancements in neural audio codec (NAC) unlock new potential in audio signal processing. Studies have increasingly explored leveraging the latent features of NAC for various speech signal processing tasks. This paper introduces the first approach to speech bandwidth extension (BWE) that utilizes the discrete features obtained from NAC. By restoring high-frequency details within highly compressed discrete tokens, this approach enhances speech intelligibility and naturalness. Based on Vector Quantized Diffusion, the proposed framework combines the strengths of advanced NAC, diffusion models, and Mamba-2 to reconstruct high-frequency speech components. Extensive experiments demonstrate that this method exhibits superior performance across both log-spectral distance and ViSQOL, significantly improving speech quality. 
\end{abstract}


\section{Introduction}
 
Speech bandwidth extension (BWE), also known as speech super-resolution, is a task in speech signal processing that aims to supplement high-frequency components to low resolution speech, enhancing speech quality and increasing its naturalness.

With the advent of deep learning, generative model-based BWE approaches have gained significant attention\cite{lee2021nu,han2022nu,mandel2023aero,yu2024bae,liu2022voicefixer,liu2024audiosr}. Various deep learning methods have been proposed to address the BWE problem including waveform-based approaches, complex spectrogram-based approaches, Mel spectrogram-based approaches, and so on. For instance, models like NU-Wave \cite{lee2021nu} and NU-Wave2 \cite{han2022nu}, based on the Denoising Diffusion Probabilistic Model (DDPM) \cite{ho2020denoising}, generate missing high-frequency contents in the time domain. In contrast, complex spectrogram-based methods such as AERO \cite{mandel2023aero} 
rely on Generative Adversarial Networks (GANs) \cite{goodfellow2014gan} to reconstruct spectrograms. 

Recently, Mel spectrogram-based methods like NVSR \cite{liu2022neural},  VoiceFixer \cite{liu2022voicefixer}, and  AudioSR \cite{liu2024audiosr} have shown promise in BWE tasks. 
All three Mel spectrogram-based methods employ HiFi-GAN \cite{kong2020hifi}, a GAN-based vocoder, to convert the processed Mel spectrograms back to the time domain. For predicting Mel spectrograms, VoiceFixer employs a GAN-based approach, while AudioSR leverages DDPM in its framework. Although performing BWE on the Mel spectrogram feature simplifies the problem, it inevitably causes information loss, particularly in reconstructing high-frequency details. This is because the Mel spectrogram compresses the entire signal, and the Mel filters cause more severe loss in high-frequency information. Although neural vocoders attempt to restore this information, the resulting speech often contains artifacts in high frequencies.

To overcome the limitation of previous Mel spectrogram-based strategies, neural network codec (NAC) \cite{defossez2022high,yang2023hifi,kumar2024high} offers a new perspective.
The proposed Vector Quantized Diffusion (VQ-Diffusion) [\citenum{gu2022vector},\citenum{tang2022improved}] 
BWE model integrates the strengths of NAC, Discrete Denoising Diffusion Probabilistic Model (D3PM) \cite{austin2021structured} and Mamba-2 \cite{gu2021efficiently,gu2023mamba,dao2024transformers}. 
D3PM provides high-quality reconstruction of discrete data,
while NAC compresses speech signals into discrete codes, achieving high compression rates and preserving speech quality. In particular, Mamba-2, a recently proposed sequence processing model builds on the state space model \cite{gu2021efficiently} and Mamba \cite{gu2023mamba}, efficiently handles sequential data and achieves excellent results. 
A key advantage of our approach is the compact and structured nature of the discrete token space, which benefits for more efficient and accurate modeling of speech.

We propose a novel VQ-Diffusion based speech BWE model (VQDiffusion-BWE). The key contributions are:

\begin{itemize} 
\item Proposing the first method to address the BWE task by leveraging D3PM to enhance performance.
\vspace{0.1cm}
\item Utilizing the generative power of VQ-Diffusion and Mamba models, combining the data compression advantages of NAC.
\vspace{0.1cm}
\item Validating the proposed method through comprehensive comparative and ablation experiments, demonstrating its superiority in both objective and subjective metrics.
\end{itemize}

\begin{figure}[htbp]
   
    \centering
    \includegraphics[width=0.49\textwidth]{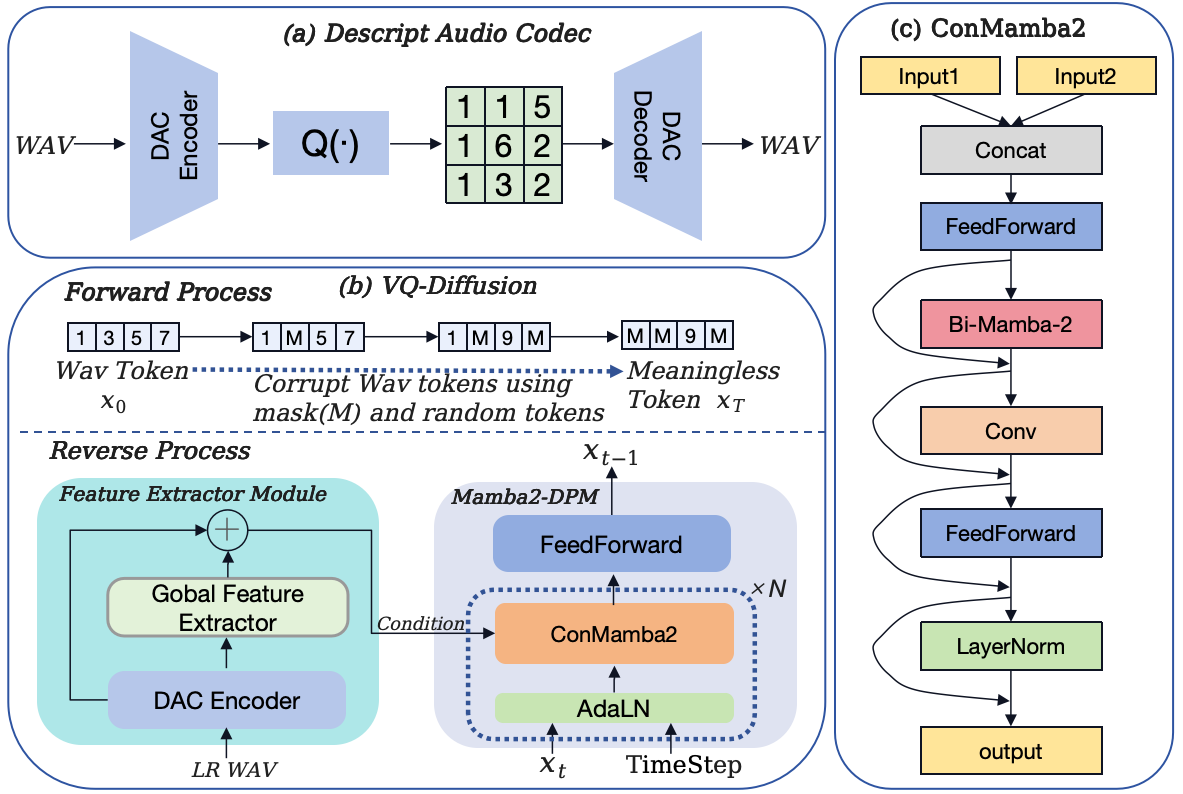} 
   \caption{Overview of the proposed model. (a) The structure of Descript Audio Codec, (b) The VQ-Diffusion process, where the top part represents the forward process and the bottom part shows the step of estimating the previous $x_{t-1}$ through a neural network, and (c) The structure of proposed ConMamba2.}
    
    \label{fig:wide_image}
\end{figure}

\section{Method}
In this section, we present the details of the proposed VQDiffusion-BWE, whose architecture overview is shown in Fig.~\ref{fig:wide_image}. VQDiffusion-BWE consists of two parts, including a feature extract module, and a Mamba-2 diffusion probabilistic model (Mamba2-DPM). For feature extract module, we use pretrained Descript Audio Codec (DAC) \cite{kumar2024high} to compress the speech data into discrete tokens. Then, extracts features from the input discrete tokens. Mamba2-DPM is used to predict the discrete token sequence conditioned on the features extracted by the previous module. Finally, the discrete token sequence is decoded into the waveform by the DAC decoder. 

\subsection{Descript Audio Codec (DAC)}\label{AA}
 
In the proposed VQDiffusion-BWE model, DAC \cite{kumar2024high} is utilized for its distinct advantages. DAC is a recently open-sourced neural audio codec. Like most NACs, it consists of three main components: an encoder, a Residual Vector Quantization (RVQ) module, and a decoder, as shown in Fig.~\ref{fig:wide_image} (a). The pre-trained model of DAC not only delivers higher fidelity in 48 kHz speech compared to other models, but also performs exceptionally well on the input speech we use for bandwidth extension (BWE), specifically those that have been processed by low-pass filtering, which removes high-frequency components. The reconstructed speech from these low-pass filtered signals does not exhibit leakage into high frequencies, the spectrogram remains sharp, and the harmonics stay intact. Those characteristics makes DAC particularly well-suited for our VQDiffusion-BWE model. Given a  random speech \( X \in \mathbb{R}^{d \cdot f_s} \) with \( d \) being the speech duration and \( f_s \) being the sample rate, DAC compresses it into \( Q \in \{1, \ldots, M\}^{d \cdot f_r \times K} \), with $f_r$ being the frame rate and  \( K \) being the number of codebooks used in RVQ and \( M \) being the codebook size.
 
\subsection{VQDiffusion-BWE}
 
The proposed VQDiffusion-BWE model integrates the strengths of DAC and D3PM. It begins by using DAC to convert data into discrete tokens. These tokens are then processed by the D3PM. D3PM extends the DDPM framework to discrete data by replacing the Gaussian noise addition operation in DDPM with a state transition approach, which is more suitable for discrete tokens. Despite handling different data types, discrete for D3PM and continuous for DDPM, the two models share a similar structure, consisting of a forward process and a reverse process. The structure of D3PM is shown in Fig.~\ref{fig:wide_image} (b), which illustrates the key components involved in the state transition approach for handling discrete tokens.

\subsubsection{Forward Process}
In the forward process, D3PM gradually corrupts the discrete tokens $x_0$  through state transitions at each step until the T-th step, resulting in a highly corrupted state $x_T$. The data at the intermediate step $t$ after state transitions is denoted as $x_t$. D3PM uses a state transition matrix to convert tokens into other tokens or mask tokens based on certain probabilities. The mask token is an additional token specifically defined to represent the absence of any information. Each token has a probability  \(\gamma_t\) to transition to a [MASK] token, a probability \(K\beta_t\) to be resampled uniformly over all \(K\) categories, and a probability of \(\alpha_t = 1 - K\beta_t - \gamma_t\) to stay the same token. The [MASK] tokens always keep fixed. The transition matrix \(Q_t \in \mathbb{R}^{(K+1) \times (K+1)}\) are defined as follows:

\begin{equation}
Q_t = \begin{bmatrix}
\alpha_t + \beta_t & \beta_t & \beta_t & \cdots & 0 \\
\beta_t & \alpha_t + \beta_t & \beta_t & \cdots & 0 \\
\vdots & \vdots & \vdots & \ddots & \vdots \\
\gamma_t & \gamma_t & \gamma_t & \cdots & 1
\end{bmatrix} .\label{eq1}
\end{equation}
The transition probability $q(x_t | x_{t-1})$ from \( x_{t-1} \) to \( x_t \) is given by:
\begin{equation}
q(x_t | x_{t-1}) = \mathbf{v}^\top(x_t) Q_t \mathbf{v}(x_{t-1}), \label{eq2} 
\end{equation}
where \(\mathbf{v}(x)\) is a one-hot column vector of length \(K\) with the entry at position \(x\) equal to 1. The categorical distribution over \( x_t \) is determined by the vector \( Q_t \mathbf{v}(x_{t-1}) \).

\subsubsection{Reverse Process}
 
The Mamba2-DPM is trained to recover corrupted token sequences using tokens extracted from the low-resolution input speech as conditions to reverse the diffusion process. The network \( p_\theta(x_{t-1} | x_t, y) \) is trained to estimate the posterior transition distribution \( q(x_{t-1} | x_t, x_0) \), where y represents the input conditions. The optimization objective is to minimize the variational lower bound.

\begin{table*}[h]
\centering
\caption{A comparison of LSD and ViSQOL scores with 48 kHz target.}
\begin{tabular}{lcccccccccc}
\toprule
\multirow{2}{*}{Method} & \multicolumn{2}{c}{4 kHz $\rightarrow$ 48 kHz} & \multicolumn{2}{c}{8 kHz $\rightarrow$ 48 kHz} & \multicolumn{2}{c}{12 kHz $\rightarrow$ 48 kHz} & \multicolumn{2}{c}{16 kHz $\rightarrow$ 48 kHz} & \multicolumn{2}{c}{24 kHz $\rightarrow$ 48 kHz}          \\
\cmidrule(lr){2-3} \cmidrule(lr){4-5} \cmidrule(lr){6-7} \cmidrule(lr){8-9} \cmidrule(lr){10-11}
  
 & LSD$\downarrow$ & ViSQOL$\uparrow$ & LSD$\downarrow$ & ViSQOL$\uparrow$ & LSD$\downarrow$ & ViSQOL$\uparrow$ & LSD$\downarrow$ & ViSQOL$\uparrow$ & LSD$\downarrow$ & ViSQOL$\uparrow$ \\
\midrule
Input      & 4.78 & 1.58 & 4.47 & 1.56 & 4.20 & 1.58 & 3.90 & 1.80  & 2.77 & 2.97\\
NU-Wave2   & 1.48 & 1.81 & 1.14 & 2.07 & 1.01 & 2.85 & 0.93  & 2.99  & 0.78  & 3.45  \\
NVSR       & 1.22 & \textbf{ 2.61 } & 1.12  & 2.79 & 1.06 & 2.85 & 1.00 & 3.01    & 0.93  & 3.50 \\
AudioSR    & 1.52  & 2.60 & 1.42 &   \textbf{ 2.79 }    & 1.30 &  2.87 & 1.22 & 2.98   & 1.07  & 3.45 \\
VoiceFixer & 1.25 & 2.44 & 1.21 & 2.56 & 1.17 & 2.61 & 1.15  & 2.73   & 1.10  & 3.07 \\
\textbf{Proposed}    & \textbf{ 1.18 }  & 2.41 &   \textbf{ 1.05 } &    2.67  &   \textbf{ 1.00 }  & \textbf{ 2.91 }  & \textbf{ 0.89 } & \textbf{ 3.15 }   & \textbf{ 0.75 } & \textbf{  3.67 }  \\
\bottomrule
\end{tabular}
\label{tab1}
 
\end{table*}

\subsection{Network}

To effectively process the discrete token for the BWE task, the proposed network is carefully designed to capture both local and global dependencies, while efficiently handling speech inputs with varying sampling rates. This design involves two main components: the feature extraction module and the Mamba2-DPM. The lower part of Fig.~\ref{fig:wide_image} (b) illustrates the process of estimating the previous step latent features \( x_{t-1} \) using the network based on timestep, condition, and latent features $x_t$ at timestep t.

\subsubsection{Feature Extraction Module}
\label{sec:feature_extraction}
The feature extraction module converts the input low sampling
rate speech into discrete tokens using DAC. Each discrete tokens extracted by DAC contain a portion of the frequency information across
all frequency bins within a time frame. Therefore, the low-pass filtering range is not prominently reflected in the token features for input speech data processed by random low-pass filtering. 
Therefore, a global feature is extracted to enhance the feature
quality. These tokens are processed by a two-layer Transformer
Encoder \cite{vaswani2017attention}, with the classification token [\texttt{CLS}] output serving as the global feature \cite{devlin2018bert}, which is then added back to the tokens. The extracted feature is used as the condition for our diffusion model, guiding the generation process.

\subsubsection{Mamba2-DPM}
\label{sec:MambaDPM}

The proposed Mamba2-DPM is composed of stacked layers of the Mamba2-DPM block, as illustrated  on the 
Fig.~\ref{fig:wide_image} (b).
The low sampling rate speech features are used as conditional inputs for subsequent model components. The first step involves applying Adaptive Layer Normalization (AdaLN) \cite{ba2016layer}, which integrates the features of $t$ and  $x_t$. AdaLN($x_t$, t) = $a_t$LayerNorm($x_t$) + $b_t$ , $a_t$ and $b_t$ are obtained from a linear projection of the timestep embedding.
 
 Once the features are normalized, latent feature $x_t$ and condition are concatenated along the feature dimension. The concatenated features are then processed through a ConMamba2 block. The ConMamba2 block, as shown in Fig.~\ref{fig:wide_image} (c), includes sub-layers such as FeedForward layer, bidirectional Mamba-2 layer (Bi-Mamba-2), convolutional layer (Conv), and Layer Normalization (LayerNorm), which work together to capture both short term and long term dependencies. The residual structure of this block helps preserve and complement the original feature information. Additionally, the Bi-Mamba-2 layer ensures that the model effectively captures both past and future information. The key component of ConMamba2 is Mamba2, which efficiently models long-range dependencies by employing a state to represent past sequences, rather than attending to the entire sequence at each time step. While reducing memory consumption, it still achieves robust performance. The effectiveness of the Mamba model in speech tasks has also been validated in multiple studies \cite{jiang2024speech,miyazaki2024exploring,zhang2024mamba}.

\section{Experiments}
\label{sec:typestyle}

\subsection{Data Configuration}
\label{ssec:subhead} 
 
We employ the VCTK dataset \cite{yamagishi2019cstr} for training, which comprises approximately 44 hours of speech from 110 speakers. To align with the data preparation strategy used in previous speech BWE studies 
[\citenum{han2022nu},\citenum{mandel2023aero},\citenum{liu2022voicefixer},\citenum{liu2022neural}]
, we omit speakers p280 and p315 due to technical problems. Consequently, the remaining 108 speakers are divided such that the last 8 are allocated for testing and the first 100 are used for training. To create inputs with missing high-frequency information, we apply low-pass filtering. The target sampling rate is set at 48 kHz, and the training input sampling rates are randomly chosen from the range of 4 kHz to 24 kHz.
 
 \subsection{Implementation Details}
\label{ssec:subhead}
 
The VQDiffusion-BWE utilizes 10 layers of Mamba2-DPM Block, each with a feature dimension of 512. A two-layer transformer, with a feature dimension of 256, is employed to extract global features. Additionally, we use 24 kHz pretrained DAC with a codebook size K of 1024 and a codebook number M of 32.
 For the VQ-diffusion, we set timesteps T = 100. The transition matrix parameters are set with \(\gamma_t\) increasing linearly from 0 to 0.9, and 
 \(\beta_t\) increasing from 0 to 0.1. The network is trained for 200 epochs and optimized using the Adam optimizer \cite{kingma2014adam}, starting with learning rate of  \(3 \times 10^{-5}\).  If the model fails to converge for two consecutive epochs, the learning rate is decayed by 0.8.

\subsection{Evaluation Metrics}
\label{ssec:subhead}
 
We evaluate the performance of our model using two metrics: log-spectral distance (LSD) and Virtual Speech Quality Objective Listener (ViSQOL) \cite{hines2015visqol}. 
These metrics comprehensively evaluate the quality of the reconstructed speech, ensuring that our model delivers high-quality results across different aspects of speech signal reconstruction.
 
\subsubsection{Log-Spectral Distance (LSD)}
 
The Log-Spectral Distance measures the similarity between the spectrograms of the original and the reconstructed signals. It is defined as follows:
\begin{equation}
\text{LSD} = \sqrt{\frac{1}{N} \sum_{n=1}^{N} \left( \log S(n) - \log \hat{S}(n) \right)^2 } \hspace{0.2cm},  \label{eq5}
\end{equation}
 
where \( S(n) \) and \( \hat{S}(n) \) represent the magnitudes of the original and reconstructed spectrograms at the \( n \)-th frequency bin, respectively, and \( N \) is the number of frequency bins.

\subsubsection{Virtual Speech Quality Objective Listener (ViSQOL)}
 
ViSQOL \cite{hines2015visqol} is a perceptual metric that estimates the quality of the reconstructed speech by comparing it to the original speech. It models human auditory perception to provide an objective measure of speech quality. The ViSQOL score ranges from 1 to 5, higher scores indicate better quality.

\section{RESULTS AND ANALYSIS}
\label{sec:majhead}
\subsection{Comparison Experiment}
\label{ssec:subhead}
\begin{figure}[htbp]
   
    \centering
    \includegraphics[width=0.49\textwidth]{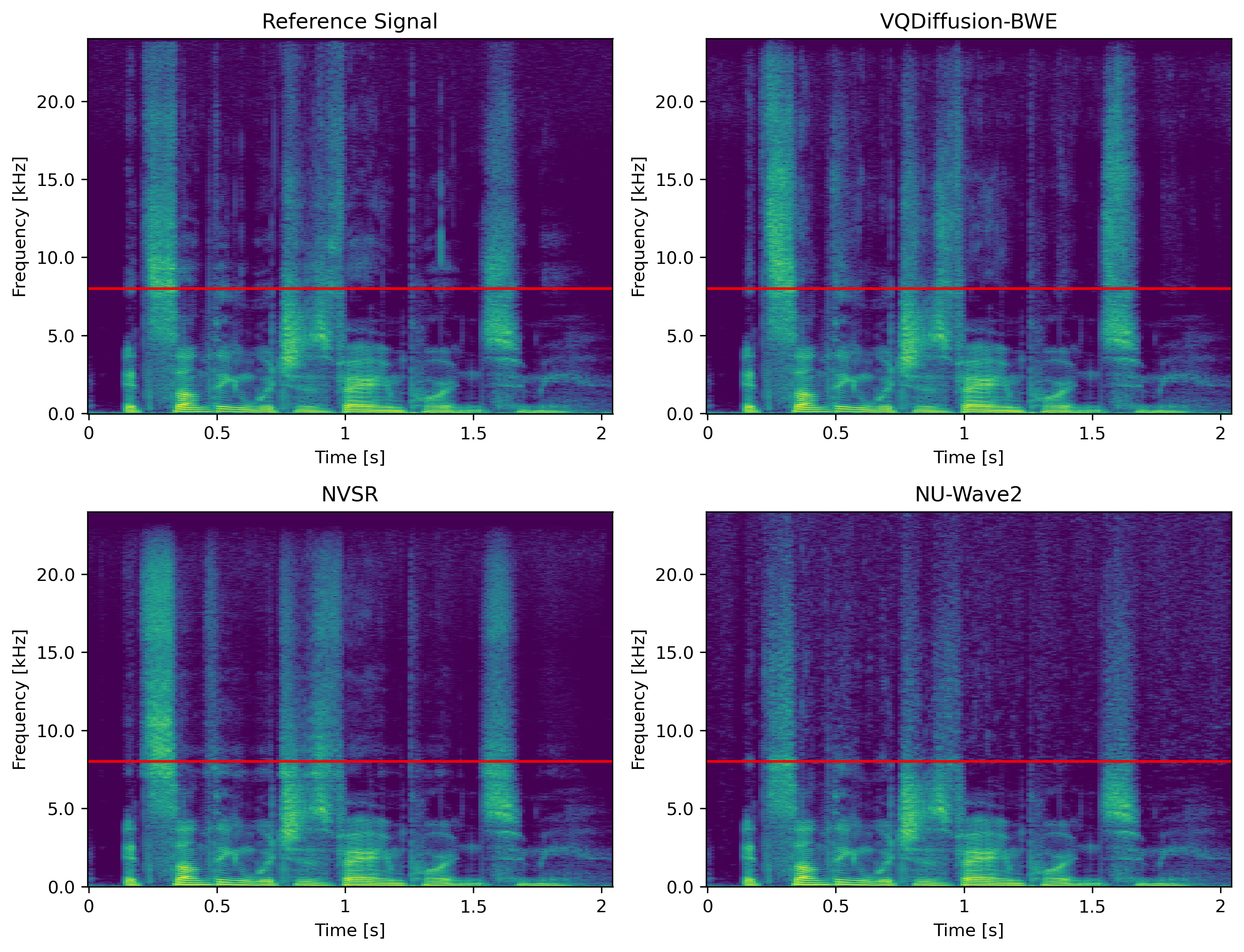} 
   \caption{Visualization of spectorgrams of reference and upsampled speeches (p360\_033) for 16 kHz input. Red lines indicate the Nyquist frequencies of downsampled signals.}
    \label{figsample}
\end{figure}

The proposed method's performance is compared to several state-of-the-art baselines, including NU-Wave2, NVSR, AudioSR, and VoiceFixer across five upsampling scenarios: 4-48 kHz, 8-48 kHz, 12-48 kHz, 16-48 kHz, and 24-48 kHz. 
This allows for a comparison of performance in generating 48 kHz target speech. As shown in Table~\ref{tab1}, the proposed model consistently outperforms others in LSD metrics. 
For ViSQOL metrics, the proposed model outperforms the baselines in the 12-48 kHz, 16-48 kHz, and 24-48 kHz tasks, but in the 4-48 kHz and 8-48 kHz tasks, the ViSQOL is slightly lower than NVSR and AudioSR.
To provide visual evidence, we present spectrograms of the reference signal and the upsampled outputs of VQDiffusion-BWE, NU-Wave2, and NVSR in Fig.~\ref{figsample}, where it can be observed that our model effectively reconstructs high frequencies, while NU-Wave2 introduces background noise and NVSR's output is smoothed and lacks fine details. 

Additionally, we downsample the model's output to obtain 24 kHz output speech. Compare the 24 kHz output with other models in the 12-24 kHz and 16-24 kHz tasks, as shown in Table~\ref{tab2}. In both tasks, our model outperforms previous baseline models in terms of LSD and ViSQOL metrics. 
Comparing the 12-24 kHz and 16-24 kHz tasks, our model performs similarly to VoiceFixer in the 12-24 kHz task but achieves a better LSD in the 12-48 kHz task. This indicates that our model generates high-frequency content more effectively compared to Mel spectrogram-based models. 

\begin{table}[t]
\centering
\caption{A comparison of LSD and ViSQOL scores with 24 kHz target.}
\resizebox{\columnwidth}{!}{
\begin{tabular}{lcccccc}
\toprule
\multirow{2}{*}{Method}  & \multicolumn{2}{c}{12 kHz $\rightarrow$ 24 kHz} & \multicolumn{2}{c}{16 kHz $\rightarrow$ 24 kHz} \\
\cmidrule(lr){2-3} \cmidrule(lr){4-5}  
   & LSD$\downarrow$ & ViSQOL$\uparrow$ & LSD$\downarrow$ & ViSQOL$\uparrow$ \\
\midrule
Input                   & 2.58 & 2.58 & 2.06 & 2.86 \\
NU-Wave2    & 1.18 & 3.12 & 1.04 & 3.20 \\
AERO        & 0.77 & 3.65 & - & - \\
NVSR       & 0.83 & 3.62 & 0.69 & 3.78 \\
VoiceFixer   & 0.78 & 3.67 & 0.63 & 3.82 \\
\textbf{Proposed}  & \textbf{0.77} & \textbf{3.75} & \textbf{0.59} & \textbf{3.87} \\
\bottomrule
\end{tabular}
}
\label{tab2}
 
\end{table}

\subsection{Ablation Study}
\label{sssec:subsubhead}
 
In the ablation experiments, we explore the effect of each step on the model. Based on the network structure, we design the following experiments.

\begin{itemize}
    
\item Ablation-1: This method replaces the DAC with HiFiCodec \cite{yang2023hifi}. The objective is to assess the impact of NAC on model performance.
\vspace{0.2cm}
\item Ablation-2: This ablation method involves removing D3PM and having the network directly predict the logits of the target discrete code, using both a frequency-domain mean absolute error (MAE) loss and a cross-entropy (CE) loss. This approach aims to investigate the effects of D3PM on the model's generative capability.

\item Ablation-3: The global feature extractor is omitted. Its purpose is to examine the influence of the global information.

\end{itemize}

The results of the ablation experiments are presented in
Table~\ref{tab3}. 
It is observed that the proposed model consistently outperforms other configurations. Additionally, each module significantly contributes to the overall performance. Specifically, for the LSD metric, the facilitating effect of NAC is 0.22, D3PM is 0.63, and the global feature extractor is 0.08. These ablation experiments validate the effectiveness of the proposed VQDiffusion-BWE framework.

     \begin{table}[H]
        \centering
        \caption{Ablation Study Results.}
         \setlength{\tabcolsep}{15pt} 
        \begin{tabular}{lcccc}
        \toprule
        \multirow{2}{*}{Method}  & \multicolumn{2}{c}{12 kHz $\rightarrow$ 24 kHz}  \\
\cmidrule(lr){2-3} \cmidrule(lr){4-5}  
   & LSD$\downarrow$ & ViSQOL$\uparrow$  \\
        \midrule
        1. Unprocessed    & 3.07  & 2.55 \\
        2. Ablation-1 & 1.15 & 3.30 \\
        3. Ablation-2 & 1.56 & 3.12\\ 
        4. Ablation-3 & 1.01 & 3.42\\ 
        5. Proposed & \textbf{0.93} & \textbf{3.55} \\ 
        \bottomrule
        \end{tabular}
        \label{tab3}
        \end{table}

\section{Conclusion}
\label{sec:print}
 
This paper introduces a novel and effective VQ-Diffusion based BWE model, which leverages NAC, D3PM, and Mamba-based neural network. The proposed method demonstrates excellent performance in BWE tasks. 
This work points to new directions for future research in BWE tasks, highlighting the potential of combining advanced neural codecs with diffusion models.

\bibliographystyle{IEEEbib}


\end{document}